\documentclass[prc,twocolumn,secnumroman,tightenlines,amssymb,aps,nobibnotes,
               superscriptaddress,showpacs,balancelastpage,floatfix,nofootinbib,natbib]{revtex4}
\usepackage{ulem}
\usepackage{graphicx}
\usepackage{multirow}           
\usepackage{textcomp}           
\usepackage{color}	
\expandafter\ifx\csname package@font\endcsname\relax\else
\expandafter\expandafter
\expandafter\usepackage
\expandafter{\csname package@font\endcsname}
\fi
\DeclareRobustCommand\substyle{\name@idx{document substyle}}
\DeclareRobustCommand\classoption{\name@idx{document class option}}
\DeclareRobustCommand\classname{\name@idx{document class}}
\def\name@idx#1#2{{\ttfamily#2}
\index{#2\space#1=\string\ttt{#2}\space#1}\index{#1>#2=\string\ttt{#2}}}

\newcommand{\jpw}[1]{{#1}}

\newcommand{\jdf}[1]{{#1}}
\newcommand{\jdff}[1]{#1}

\renewcommand{\bf}[1]{\begingroup\bfseries\mathversion{bold}#1\endgroup}

\begin{document}
\title[] {Sequential fission of highly excited compound nuclei in a 4D Langevin approach
          }
       
\author{D. Gruyer}
\email{gruyer@lpccaen.in2p3.fr}
\affiliation{LPC, IN2P3-CNRS, ENSICAEN et Universit\'e de Caen, F-14050 Caen Cedex, France}
\author{K. Mazurek}
\email{katarzyna.mazurek@ifj.edu.pl}
\affiliation{Institute of Nuclear Physics Polish Academy of Sciences, PL-31342 Krakow, Poland}
\author{J. D. Frankland}
%\email{frankland@ganil.fr}
\affiliation{Grand Acc\'el\'erateur National d'Ions Lourds, CEA/DSM-CNRS/IN2P3, Bvd Henri Becquerel, 14076 Caen, France}
\author{E. Bonnet}
%\email{bonnet@ganil.fr}
\affiliation{Laboratoire SUBATECH, 4 Avenue A. Kastler, F-44 072 Nantes Cedex 03, France}
\affiliation{Grand Acc\'el\'erateur National d'Ions Lourds, CEA/DSM-CNRS/IN2P3, Bvd Henri Becquerel, 14076 Caen, France}
\author{P. N. Nadtochy}
%\email{nadtoch77@yahoo.com}
\affiliation{Omsk State Technical University, Mira prospekt 11, Omsk, 644050, Russia}
\author{A. Chbihi}
%\email{chbihi@ganil.fr}
\affiliation{Grand Acc\'el\'erateur National d'Ions Lourds, CEA/DSM-CNRS/IN2P3, Bvd Henri Becquerel, 14076 Caen, France}
\author{J.P. Wieleczko}
\affiliation{Grand Acc\'el\'erateur National d'Ions Lourds, CEA/DSM-CNRS/IN2P3, Bvd Henri Becquerel, 14076 Caen, France}

\date{\today}

%\maketitle

\begin{abstract}

\jdf{In highly dissipative collisions between heavy ions,}\ the optimal conditions to investigate \jdf{different} de-excitation channels \jdf{of hot nuclei} such as evaporation, fission or multifragmentation are well known. \jdf{One} crucial issue \jdf{remains} the excitation energy region where fission \jdf{gives way to} multifragmentation. \jdf{In this paper,}\ the onset of \jdf{multi-fragment exit channels} is investigated \jdf{in terms of} sequential fission. \jdf{For the first time,}\ the dynamical approach based on solving \jdf{Langevin transport equations} in multidimensional collective coordinate space is used to \jdf{follow} the de-excitation of highly excited (up to $E^{\star}$=223--656~MeV) \jdf{$^{248}$Rf compound nuclei}. \jdf{The sequential fission model we propose} contains two steps: (1) time evolution of the compound nucleus \jdf{up to either scission or residue formation,}\ followed by (2) dynamical calculations of each primary fragment separately. This procedure allows to obtain from one to four cold fragments correlated with the \jdf{light}\ particles emitted during the de-excitation process.  Experimental data measured with the INDRA detector for the $^{129}$Xe+$^{nat}$Sn reaction at beam energies 8, 12 and 15~MeV/nucleon \jdf{provide strong constraints for this}\ sequential fission scenario.
\end{abstract}
\maketitle

\section{Introduction}

The nature and magnitude of nuclear dissipation is one of the most interesting and challenging problems in nuclear
dynamics. It has been extensively studied through large-amplitude collective motions in nuclei such as fission of compound nuclei at moderate excitation energy (see \cite{mazurek:2017} and refencences therein):
dissipation directly influences fission timescales and fission fragment properties.
The \jdf{nuclear}\ potential energy surface \jdf{(PES)}\ also plays a crucial role in low energy nuclear fission \cite{mazurek:2011,mazurek:2013,mazurek:2013a}. Fission fragment properties and light particle evaporation result from
the complex interplay between \jpw{\jdf{PES}, rotational, and dynamical effects} (viscosity and inertia). These two critical inputs of fission models are very difficult to
disentangle, so the strength and deformation dependence of the one-body viscosity is still debated. It is often parametrized using the ``wall-plus-window''  formula \cite{blocki:1978} by reducing the ``wall'' term with
a factor $k_s$ whose value can vary from $0.2$ to $0.5$ \cite{nadtochy:2014}. In addition to these constant-$k_s$ parametrizations, the authors of \cite{pal:1998}
proposed a deformation-dependent reduction coefficient $k_s(q_1)$ reflecting the degree of particle motion randomization.

Studying fission of the compound nucleus (CN) at high excitation energy would allow to partially decouple these two contributions to fission. In fact, for excitation energy much higher than the fission barrier
the system is more sensitive to dissipation energy modes than the details of the potential.
The counterpart is that increasing the excitation energy would also increase the complexity and the number of possible exit channels.
Indeed, central heavy-ion collisions well above the Coulomb barrier but \jdf{below} the Fermi energy regime \jdf{can engender}\  many processes leading to the production of one, two, three or more
heavy fragments accompanied by many light particles. This energy regime has been scanned by the INDRA collaboration \jdf{using a powerful charged particle multidetector to obtain highly exclusive data on}\  $^{129}$Xe+$^{nat}$Sn central collisions from 8 to 25~MeV/nucleon \cite{chbihi:2013,manduci:2016}.
This unique set of data allows to investigate the transition from a \jdf{fission-dominated} regime at 8~MeV/nucleon,
the emergence of sequential fission producing three or four heavy fragments \jdf{\cite{gruyer:2015}}, up to the pure multi-fragmentation regime above 20~MeV/nucleon \jdf{\cite{Gruyer2013Nuclear}}. 
This energy \jdf{range} \jpw{would appear} to be a wonderful laboratory to study nuclear viscosity with increasing temperature while minimizing the influence of the nuclear potential energy surface.

From a theoretical point of view, no \jdf{single} dynamical model is able to describe \jdf{satisfactorily} the evolution \jdf{from the}\  fission \jdf{to the}\  multi-fragmentation \jdf{regime}:
this energy region is at the \jdf{crossroads} between two general approaches. 

On the high energy side, multi-fragmentation has been intensively studied with \jdf{statistical models \cite{Zhang1987Decay,Zhang1987Decay2,Bondorf1995Statistical,Raduta1997Simulation}}, molecular dynamics \cite{hartnack:1989,aichelin:1991,ono:1992,papa:2001}
or stochastic mean field methods \cite{danielewicz:1991,colonna:1998,napolitani:2013,shvedov:2010}. These models are usually applied for reactions above 25~MeV/nucleon beam energy and are not suitable for \jdf{a correct description of} fission.

On the low energy side, there is a large variety of models describing fission.
It can be investigated with pure statistical approaches by using codes like GEMINI++ \cite{charity:2010} or GEF (for heavy and super-heavy nuclei)
\cite{schmidt:2016}. These codes give a good estimation of light particles and  $\gamma$-ray evaporation. 
Fission fragment (FF) and evaporation residue (ER) mass/charge distributions are also well reproduced for many reactions.
However, for a study of the dissipation process during equilibration of the colliding system and its path to fission, dynamical approaches are mandatory. 
Many quantum and classical methods describing the time evolution of fissioning \jdf{nuclei} are available nowadays. 
Self-consistent models such as Time Dependent Hartree-Fock with the BCS (Bardeen-Schrieffer-Cooper) pairing \cite{washiyama:2009,goldsim:2009} or more sophisticated energy-density-based methods \cite{bulgac:2016} are well-grounded but also extremely time-consuming. 
Classical transport models based on the Metropolis walk \cite{metropolis:1953}, Smoluchowski \cite{moller:2015} or Langevin equations \cite{adeev:1988en}, combined with macroscopic-microscopic potentials, 
are more appropriate to investigate dissipation effects \cite{mazurek:2017}. 
The main advantage of this dynamical treatment is an access to the time evolution of the fissioning nucleus. Sequential fission has been already obtained by coupling such a dynamical fission model with a statistical secondary decay code \cite{karpov:2017} in order to remove sequential events. 
%\jpw{\sout{However, sequential fission has never been considered in a dynamical way up to now. }}
%\jpw{However, the two successive fission steps have never been considered in a coherent and fully dynamical way up to now. }

The main goal of this work was to develop a model able to describe sequential fission treating, \jpw{for the first time}, the two \jpw{successive} steps in a coherent and fully dynamical way. 
The article is organized as follows:
the sequential fission model is described in Section \ref{mod}. Experimental details are briefly presented and a comparison between the model predictions and some selected observables 
from INDRA data (particle multiplicities and charge distributions) are discussed in Section \ref{comp}. 
Finally, Section \ref{conclu} contains closing remarks.

\begin{figure}[tbh]
\begin{center}
\includegraphics[width=.45\textwidth]{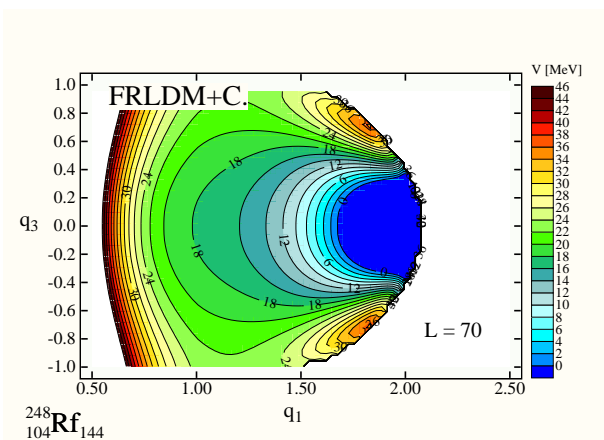}
\caption{The potential energy surface for $^{248}$Rf in the elongation-mass-asymmetry plane for spin L=70~$\hbar$.}
\label{pes248Rf}
\end{center}
\end{figure}

\section{Sequential fission model}
\label{mod}

\subsection{Fission dynamics}%Langevin equations}

The details of the fission model used in this work are presented in i.e. 
\cite{nadtochy:2012,mazurek:2013,mazurek:2017} thus here only a few remarks will be given.

The evolution of the system towards fission is obtained by solving the coupled Langevin
classical equations of motion, where the combined action of
the driving potential, friction, and diffusion forces determines the trajectory of the nucleus on the three-dimensional potential energy surface (PES).
The Langevin equations are solved in four dimensional collective coordinates (CC) space.
This collective coordinate space contains three
parameters ${\rm\vec{q}}$=(q$_1$, q$_2$, q$_3$) responsible for the deformation of the CN, based on the ``funny hills'' parametrization ($c,h,\alpha$) \cite{brack:1972} with elongation of 
the nucleus (q$_1$), its neck size (q$_2$), and mass-asymmetry constraints (q$_3$). The fourth collective coordinate is the projection of the angular momentum into the fission axis \cite{nadtochy:2012}. This degree of freedom
has been introduced to orient the CN in the laboratory frame  which gives access to fission fragment angular distributions.

\begin{figure}[tbh]
\begin{center}
\includegraphics[width=.99\linewidth]{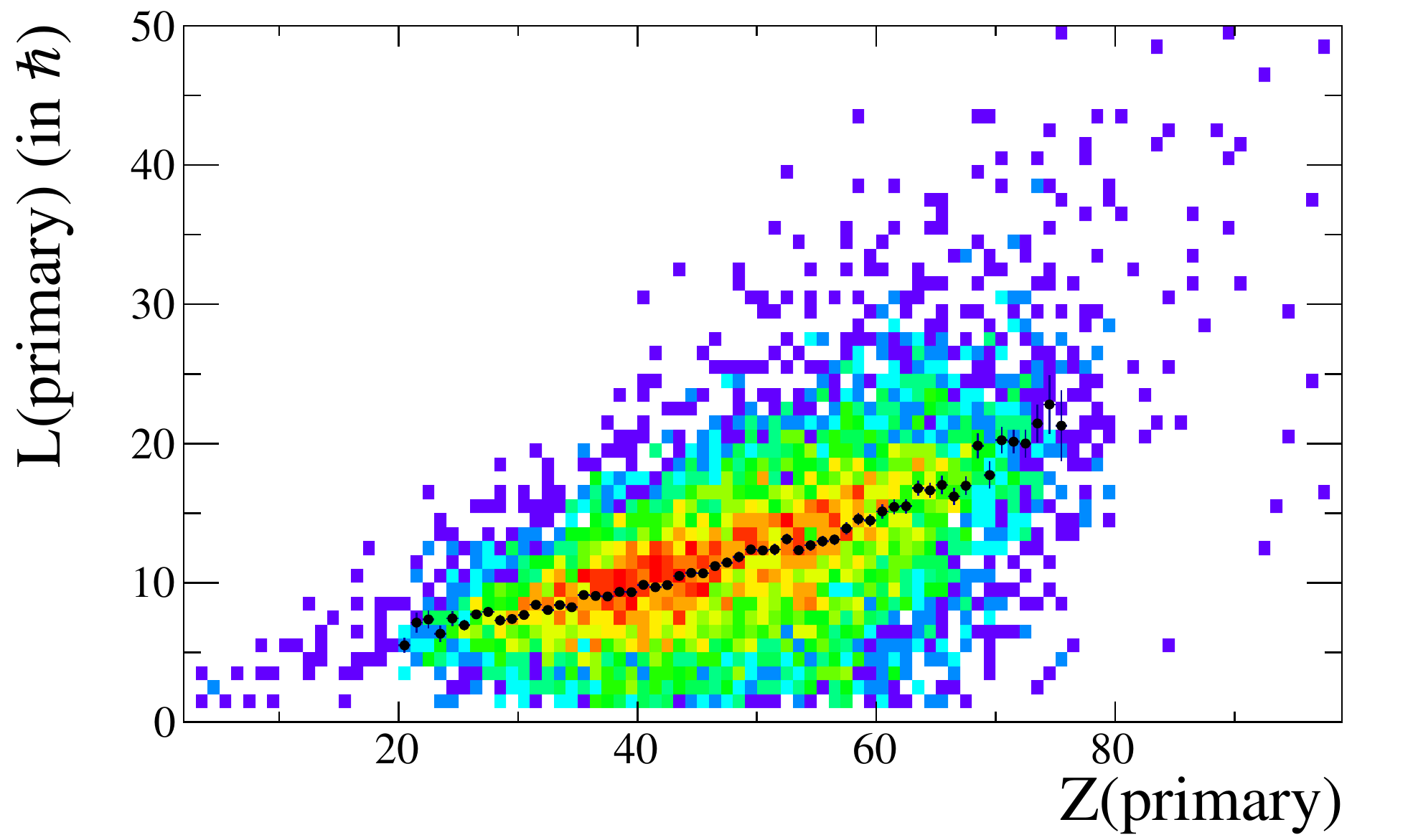}
\caption{(color online) The nucleus of $^{248}$Rf excited to energy $E^*$= 223~MeV divides into two parts (primary fission fragments) with charge and angular momentum distribution presented here. 
Full symbols give the average evolution.}\label{lz_distr}
\end{center}
\end{figure}

\begin{figure*}[tbh]
\includegraphics[width=0.89\textwidth]{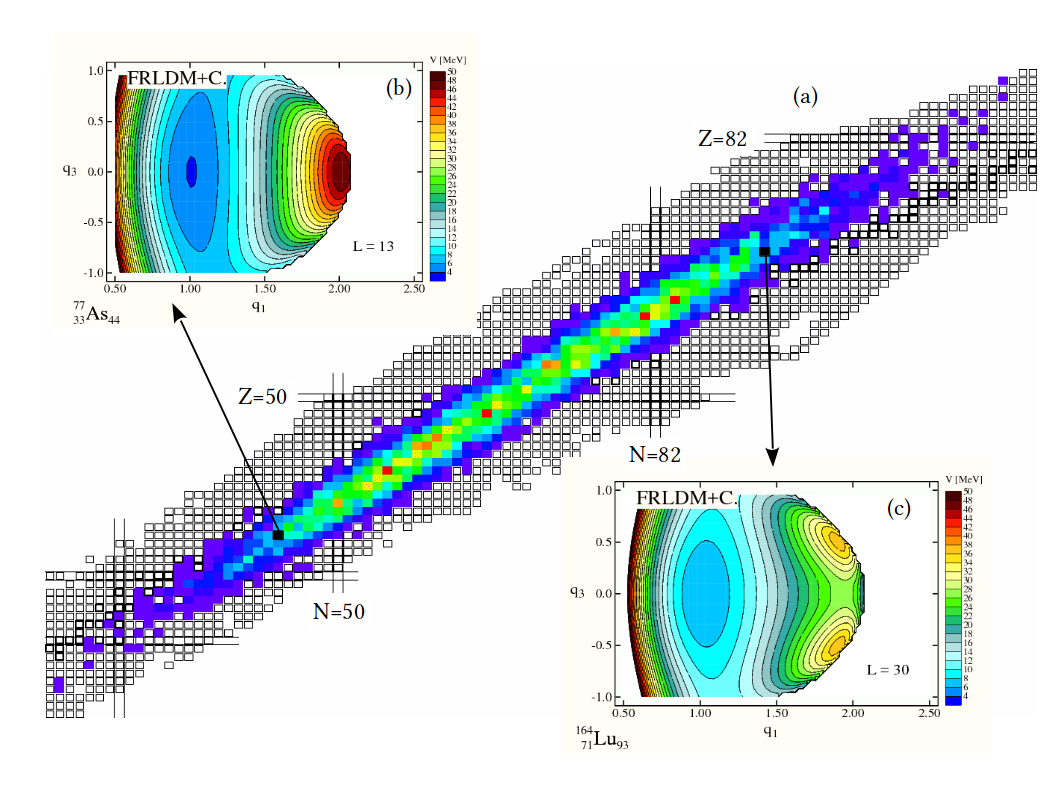}
\caption{(color online) \jdf{(a) Distribution of the primary fission fragments' proton and neutron number (vertical and horizontal scales, respectively). Color-intensity scale represents the number of FF produced for a given $(N,Z)$. In the insets are examples of} PES for two complementary FF: \jdf{(b)} $^{77}$As and \jdf{(c)} $^{164}$Lu at angular momenta: 13~$\hbar$ and 30~$\hbar$, respectively.}
\label{primary_pes}
\end{figure*}

The driving potential is the Helmholtz free energy $\rm F({\vec{q}})=V({\vec{q}}) - {\it a}({\vec{q}}) T^2$ which 
contains the PES ($V({\vec{q}})$) and the entropy related 
term $a({\vec{q}}) T^2$ where $a$ is the deformation-dependent level density parameter proposed by 
Ignatyuk \cite{ignatyuk:1975}. The temperature of the ``heat bath'' $\rm T$ has been determined by the Fermi-gas model
formula $\rm T=(E_{int}/{\it a})^{1/2}$, where $\rm E_{int}$ is the
internal excitation energy of the nucleus. 
The diffusion tensor $\theta_{ij}$ responsible for the random character of the process is derived from Einstein's relations $\rm \sum \theta_{ik}\theta_{kj} = T \gamma_{ij}$ and 
its stochastic nature is ensured by the normalized Gaussian white noise $\xi_j\left(t\right)$.
Energy dissipation is driven by the mass and the friction tensors. 
The mass tensor $m_{ij}({\vec{q}})$ is obtained from the Werner-Wheeler approximation of incompressible irrotational flow \cite{davies:1976},
while the friction tensor $\gamma_{ij}({\vec{q}})$ is derived from the ``wall-plus-window'' one-body 
dissipation mechanism \cite{blocki:1978} with the possibility of reducing the ``wall'' part by the viscosity reduction factor $k_s$ \cite{karpov:2001,nadtochy:2002,nadtochy:2012,nadtochy:2014}.
The original ``wall-and-window'' formula is recovered for $k_s$=1, while the super-fluid limit is obtained for $k_s$=0.  
The bigger $k_s$ provides more viscous system leading to longer fission time and higher 
particle multiplicity. 
The potential energy surface (PES) $\rm V({\vec{q}})$ is calculated in every point of this CC space, using the finite range liquid drop model (FRLDM) 
with the Wigner term included \cite{moller:2004}. The rotational energy is then added assuming the rigid-body regime. Fig.~\ref{pes248Rf} presents the PES of $^{248}$Rf in 
the elongation-mass-asymmetry plane for an angular momentum L=70~$\hbar$.

The initial conditions of the system are assumed to correspond to a spherical compound nucleus (q$_1$=1) with a total excitation
energy E* given by the entrance channel of the reactions.
%The angular momentum L for each Langevin trajectory is sampled from a triangular distribution function with L$_{\text{max}}$=130$\hbar$. %\jdf{why?}. 
During its path to fission the system can de-excite by evaporating light particles
(n, p, d, t, $\alpha$, and $^{3}$He) and $\gamma$-rays using a Monte Carlo approach. The decay width for the emission
of a given particle is calculated with the statistical code based on Hauser-Feschbach theory.

If the final shape is necked-in, the trajectory is marked as a fission event, otherwise it is accounted to the evaporation residue channel. 
For such a massive nucleus the macroscopic fission barrier is very small (see Fig.~\ref{pes248Rf}), thus the evaporation residue channel is \jdf{strongly inhibited}. The movement of the nucleus over the PES
is mainly \jdf{governed} by energy dissipation since the sensitivity of the system to the details of the PES is negligible. The excitation energies for the three reactions which are 
discussed here are (\jdf{calculated from the mass balance assuming complete fusion)}: $E^*$= 223, 471 and 656~MeV.
\jpw{The angular momentum L for each Langevin trajectory is sampled from a triangular distribution function with L$_{\text{max}}$=130$\hbar$.}
% spin predicted for \dg{capture} is around 130~$\hbar$ \cite{gruyer:2015}.
%\jdf{[Maximum spin for fission (disappearance of barrier) or for capture (disappearance of pocket)?]}. 
\jdf{These} conditions are at the limit of the applicability of the classical Langevin equations. 
Nevertheless after evaporation of prescission particles, the nucleus evolves to scission in the usual way and we are able to \jdf{track} the excited nucleus on every trajectory.

%In case of fission, the angular momentum and excitation energy remaining in the system is shared among fission fragments 
%considering the tilting mode \cite{schmitt:1988}. 

\jpw{In case of fission, the remaining excitation energy is shared among fission fragments considering thermal equilibrium \cite{nadtochy:2003en} while the angular momentum transfered to fission fragments is calculated using Eq.5 of \cite{schmitt:1988}.}
Fission fragments produced after this first step are now called ``primary fission fragments'' (PFF).
Fig.\ref{lz_distr} shows the correlation between primary fission fragment
charge and residual angular momentum after fission of $^{248}$Rf with $E^*$= 223~MeV excitation energy. It can be seen that on average $L_{PFF}$ increases with increasing $Z_{PFF}$ (black points on Fig.\ref{lz_distr}), but the correlation is quite large due to the initial CN angular momentum distribution and particle evaporation.

\subsection{Sequential procedure}

The distribution of primary fission fragment charge and \jdf{neutron number} is displayed in Fig.\ref{primary_pes}(a) for the fission of $^{248}$Rf with $E^*$= 223~MeV excitation energy. 
They populate the nuclear chart over a wide range of mass, N/Z ratio and fissility. % and some of them could have enough residual excitation energy and angular momentum (see Fig.\ref{lz_distr}) to potentially undergo a secondary fission.
The main question we want to address in this work is the following: \textbf{Do some of these primary fission fragments possess enough residual excitation energy and angular momentum to undergo a secondary fission?}

Fig.~\ref{primary_pes}(b$-$c) shows as an example the PES of the calculated primary FF: $^{77}$As with $L_p$=13~$\hbar$ and $^{164}$Lu with $L_p$=30~$\hbar$ angular momenta. 
These two nuclei are eventual candidates for secondary fission. The fission barrier for $^{77}$As is more than 40~MeV thus it will finish as an evaporation residue or very asymmetric secondary fission 
as the spin is too low to deexcite by particle emission. For $^{164}$Lu, the symmetric fission barrier is around 30~MeV and with angular momentum 30~$\hbar$ it can evaporate some particles or \jdf{undergo} secondary fission. 
This discussion is very simplified but it shows that to study correctly the secondary fission we have to calculate the library of all possible PES for each nucleus that
could be produced during the primary fission.

The originality of the present model is that each of the primary FF is treated as \jdf{a potential} fissioning nucleus with the same approach as for the CN primary fission. %and the 4D Langevin code presented before is applied once again.
The dynamical evolution of each primary FF is done by solving the set of Langevin equations in multidimensional collective coordinate space.
The new excited nucleus emits particles and it can finish as an evaporation residue or fission fragment, called now ``secondary fission fragments''.

To summarize, our procedure to investigate the sequential fission is composed of two parts.
Firstly, we compute the time evolution of the highly excited compound nucleus produced in heavy-ion collision which provides the mass/charge/kinetic energy of evaporation residues and primary fission fragments.
Secondly each primary FF is treated dynamically with the same fission model to deexcite as a secondary fission fragment or evaporation residue.

These two steps give access to the final multiplicity of fragments (M$_{\text{frag}}$). Four different cases can occur:
\begin{enumerate}
 \item compound nucleus ending as an evaporation residue (M$_{\text{frag}}$=1);
 \item primary fission of the CN with both fragments ending as secondary evaporation residues (M$_{\text{frag}}$=2);
 \item primary fission of the CN with one primary fragment undergoing secondary fission while the complementary ends as secondary ER (M$_{\text{frag}}$=3);
 \item primary fission of the CN with both primary fragments undergoing secondary fission (M$_{\text{frag}}$=4). 
\end{enumerate}
The full knowledge about particles emitted during each step, as well as their energies, are also available for comparison with experimental data. \jpw{ $\sim$$10^5$ trajectories were generated for the primary fission step and $\sim$$5.10^3$ complete calculations were performed for each excitation energy/viscosity combination.}

\begin{figure}[tbh]
\begin{center}
\includegraphics[width=.99\linewidth]{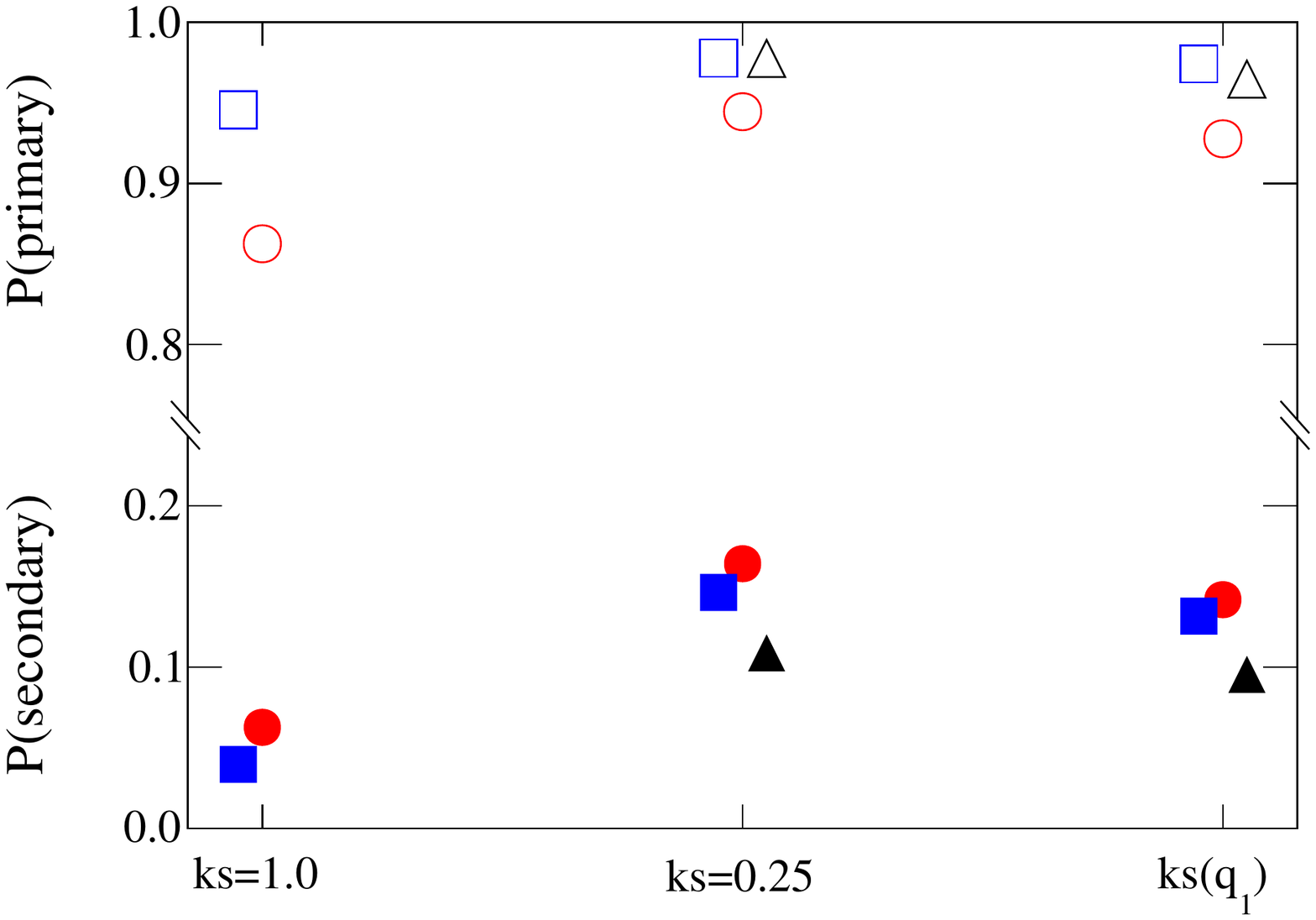}
\caption{(color online) $^{248}$Rf fission probabilities with excitation energy $E^*$=223~MeV (circles),  471~MeV (squares), and 656~MeV (triangles) for different viscosity \jpw{parametrization}s: 
primary fission probability (empty points) and secondary fission probability per primary fission fragment (full points).}\label{pprisec}
\end{center}
\end{figure}

%\begin{figure}[tbh]
%\begin{center}
%\includegraphics[width=.99\linewidth]{pprisec90.pdf}
%\caption{(color online) same as Fig. \ref{pprisec} with lmax=90$\hbar$.}
%\end{center}
%\end{figure}

Primary and secondary fission probabilities for the three excitation energies are displayed on Fig.~\ref{pprisec} as a function of the viscosity \jpw{parametrization}. 
Primary fission probabilities are slightly affected by the type and strength of dissipation: the higher the viscosity, the smaller the primary fission probability.
This effect is strongly enhanced when looking at secondary fission probabilities.
\jpw{In fact, in case of high energy dissipation the compound nucleus emits a lot of particles
during its path to fission, reducing the excitation energy and angular momentum to be
transfered to fission fragments and therefore reducing the probability of secondary fission.}
%In fact, in case of high energy dissipation the compound nucleus emits a lot of particles during its path to fission, reducing the excitation energy and angular momentum to be transfered to fission fragments. 
%However, fission fragments need excitation energy and angular momentum to undergo secondary fission whose \jdf{probability} will \jdf{therefore be}\  reduced.
It demonstrates that sequential fission at relatively high excitation energy is a very sensitive tool to probe energy dissipation while minimizing potential energy effects.
%\dg{Could be very nice to show a calculation example with ks=0.25 and an other PES (LSD?). to be discussed.}\jdf{ok if you have it}
Since the aim of the present article is to understand the mechanism of multi-fragment production \jdf{by sequential fission}, all results presented in the following are obtained with $k_s=0.25$.
The influence of the choice of viscosity parametrization on sequential fission final observables will be addressed in a \jpw{forthcoming} article.

\section{Results}
\label{comp}

In the following, model calculations will be compared with $^{129}$Xe + $^{119}$Sn collisions at 8, 12, and 15~MeV/A beam energy whose complete fusion would lead to $^{248}$Rf compound nucleus
with excitation energies $E^*$=223~MeV,  471~MeV, and 656~MeV. These data were measured with the INDRA multidetector at the GANIL accelerator facility.
INDRA \cite{pouthas:1995} was initially designed to study \jdf{multifragmentation in Fermi} energy collisions but the low identification thresholds \jdf{of $\sim 1-2$~MeV/nucleon even for heavy} charged products also allows to investigate lower energy reactions.
Experimental details can be found in \cite{gruyer:2015}.
For the relatively low energy reactions discussed here, INDRA permits
\jdf{a highly-exclusive measurement of coincident fission fragments and light charged particles for each recorded reaction}.
On the other hand, detection efficiency for \jdf{reactions leading to a heavy ER recoiling close to $0^{o}$}
is very low so this \jdf{exit} channel will not be considered here.

%The secondary fission is investigated here in the $^{129}$Xe+$^{119}$Sn reaction at three beam energies: 8, 12 and 15 MeV/nucleon. 
These \jdf{studies} were extremely difficult as the usual range of excitation energies where the fission dynamics is studied with 4D Langevin code is $E^{\star}$=50--250~MeV and 
the presented reactions lead to $E^{\star}$=223--656~MeV. At \jdf{such} high energies, the details of the potential energy surfaces are less important than \jdf{effects due to dissipation}. 
The evaporation of \jdf{light} particles before scission depends mainly on the excitation energy and the time needed to pass from the spherical initial shape to a well necked-in form of the scission surface. 
There are several observables that could be compared between existing experimental data measured with \jdf{INDRA} and theoretical calculations.
In the present article we will focus our attention on the \jdf{light} particle multiplicities and final charge distributions \jdf{for fission fragments}.

\subsection{Asymmetric primary charge partition}

A detailed analysis of experimental 3-fragment events has been presented in Ref. \cite{gruyer:2015}  
\jdff{where an even-by-event method has been proposed to identify the sequence of splitting.}
%\jpw{an event-by-envent} method to identify the sequence of splitting event by event has been proposed.
With this method, each fragment can be identified as coming from the primary or the secondary fission which allows to reconstruct experimentally the FF charge distribution after the primary fission.
The PFF distribution presents two well separated bumps, while the secondary FF is symmetric (Fig.~6 of \cite{gruyer:2015}). 
This asymmetric primary fission is absolutely not expected for such a high excitation energy and has been interpreted as an effect of selecting only 3-fragment events:
at least one PFF has to be heavy enough to possess a small fission barrier mandatory for the secondary fission.

\begin{figure}[tbh]
\begin{center}
\includegraphics[width=.99\linewidth]{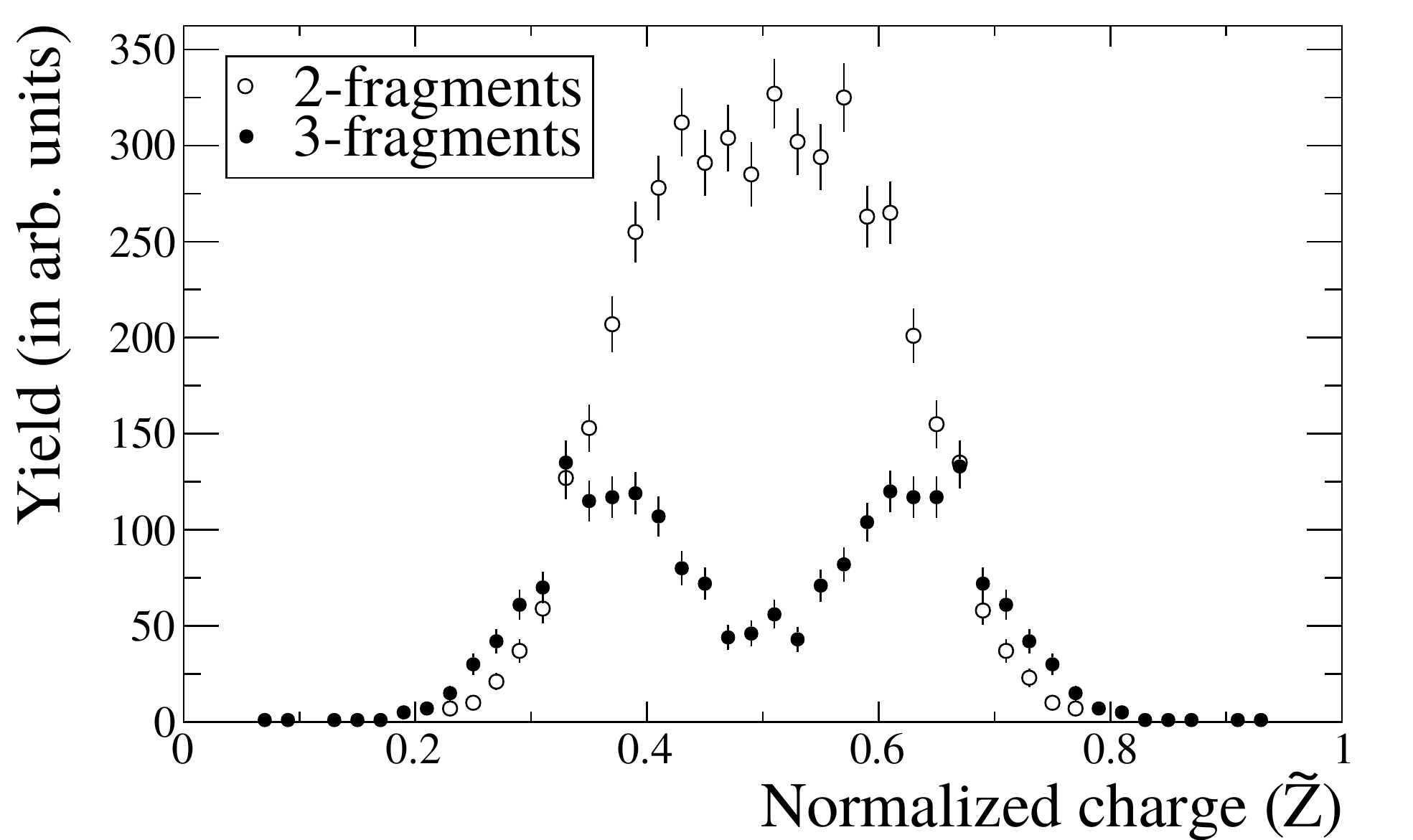}
\caption{Normalized charge ($\tilde{Z}$) distribution of the primary fission fragments for 2-fragment events (empty points) and 3-fragment events (full points) 
\jdf{calculated for} the deexcitation of $^{248}$Rf ($E^*$=223~MeV) in the 4DLangevin model.}
\label{zfragzcomp}
\end{center}
\end{figure}
The analysis of the full fission fragment charge/mass distribution displayed in Fig.~\ref{primary_pes} gives the possibility to connect the number of final fragments with its sources. Hence
 Figure~\ref{zfragzcomp} shows the distribution of normalized charge ($\tilde{Z}^i=Z_{FF}^i/ \sum Z_{FF}^i$) of primary fission fragments predicted by the model for $E^*$=223~MeV, 
considering separately events with 2-fragments and 3-fragments in the final state. 
With the present approach we recover the behavior observed experimentally: by considering 3-fragment events, we select the most asymmetric primary fission, independently of the global charge distribution. 
However, in our model the lowering of the fission barrier is mainly due to the residual angular momentum.
In such a case, the heavy fragments could undergo secondary fission if the angular momentum is high enough to lower the fission barrier. 
Since residual angular momentum increases \jdf{with PFF charge/mass (see Fig.~\ref{lz_distr})}, only the most asymmetric primary fission lead to 3-fragment events, thus confirming the interpretation proposed in \cite{gruyer:2015}. 
The fragments coming from the symmetric division of CN (open points) have mass around A=120 and angular momenta around 10-20~$\hbar$ which provide high fission barriers. 
These nuclei de-excite by particle evaporation and \jdf{secondary} fission probability is very low.

\begin{table}[!htbp]
\centering
\begin{tabular}{l l c c}
\hline
		&	&\textlangle M(Z=1)\textrangle	&\textlangle M(Z=2)\textrangle	\\
\hline
8~MeV/A~	& exp	& 1.6	&0.9	\\
%	& $k_s=1$	& 1.6	&0.5	\\
	& $k_s=0.25$	& 1.4	&0.6	\\
%	& $k_s(q_1)$	& 1.5	&0.6	\\
\hline
12~MeV/A~	& exp	& 4.1	&2.6	\\
%	& $k_s=1$	& 5.5	&1.5	\\
	& $k_s=0.25$	& 5.2	&1.6	\\
%	& $k_s(q_1)$	& 5.3	&1.6	\\
\hline
15~MeV/A~	& exp	& 5.3	&3.5	\\
	& $k_s=0.25$	& 7.5	&2.2	\\

% \multirow{2}{*}{toto} & exp  \\
%                       & theo \\                                        
\hline
\end{tabular}
\caption{\jdf{Mean multiplicities for hydrogen ($Z=1$) and helium ($Z=2$) isotopes for the experimental data at the three beam energies indicated, compared to the results of calculations for the corresponding excitation energies.}
\label{tab:multpart}}
\end{table}

\begin{figure}[tbh]
\begin{center}
\includegraphics[width=.99\linewidth]{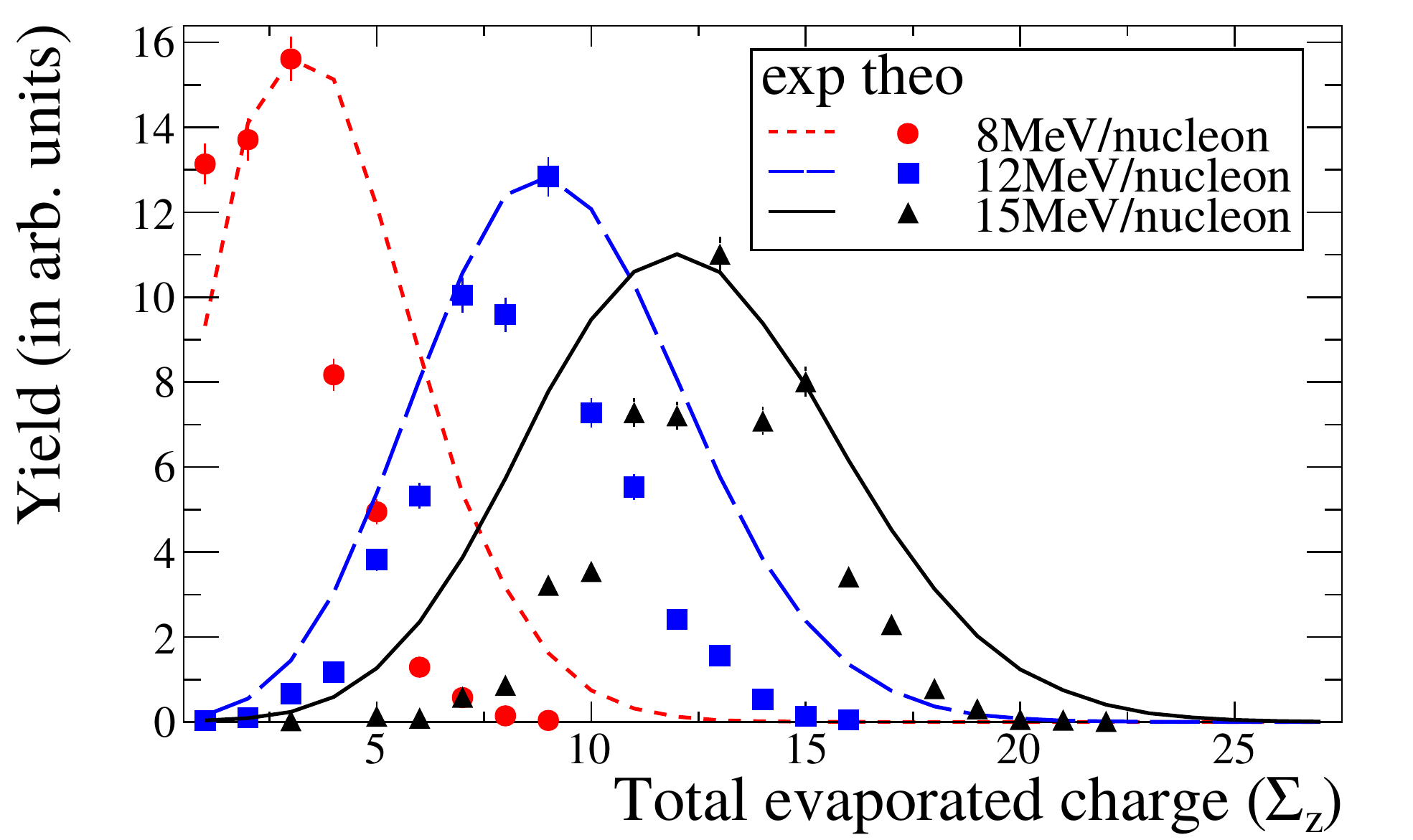}
\caption{(color online) Distribution of the sum of evaporated charges ($\Sigma_Z$) for the decay of $^{248}$Rf at three excitation energies compared to data. Simulation were performed using $k_s=0.25$.}
\label{multpart}
\end{center}
\end{figure}

\subsection{Particle multiplicities}

The \jdf{decay} of the hot \jdf{CN into} two, three or four fragments strongly depends on the \jdf{number of light particles} emitted during the process. 
Before primary scission, particles are emitted randomly by the CN during its path from initial point to the well-necked shape. Also after the division into two fragments, each of them could possess enough spin and/or 
excitation energy to deexcite by releasing some particles or $\gamma$ rays. Table~\ref{tab:multpart} shows the average multiplicity of hydrogen (Z=1) and helium (Z=2) isotopes emitted during the cooling of the excited $^{248}$Rf 
for the three \jdf{excitation} energies considered here.
%\jdf{and with different viscosity choices, compared to the experimental values for the corresponding beam energies}. 
\jdf{At the beam energy 8~MeV/nucleon ($E*=223$~MeV) the agreement is very good; for higher energies, the} model systematically overestimates the mean number of evaporated particles with charge Z=1 while 
underestimating the number of helium \jdf{species} (Z=2).
%\jdf{These multiplicities show little sensitivity to the choice of viscosity parameter, $k_s$.}
However, \jdf{as shown in Fig.~\ref{multpart},} the distribution of total evaporated charge ($\Sigma_Z=\sum_{Z<3}{(Z)}$) is qualitatively consistent with experimental data.

\begin{figure*}[tbh]
\begin{center}
\includegraphics[width=.32\textwidth]{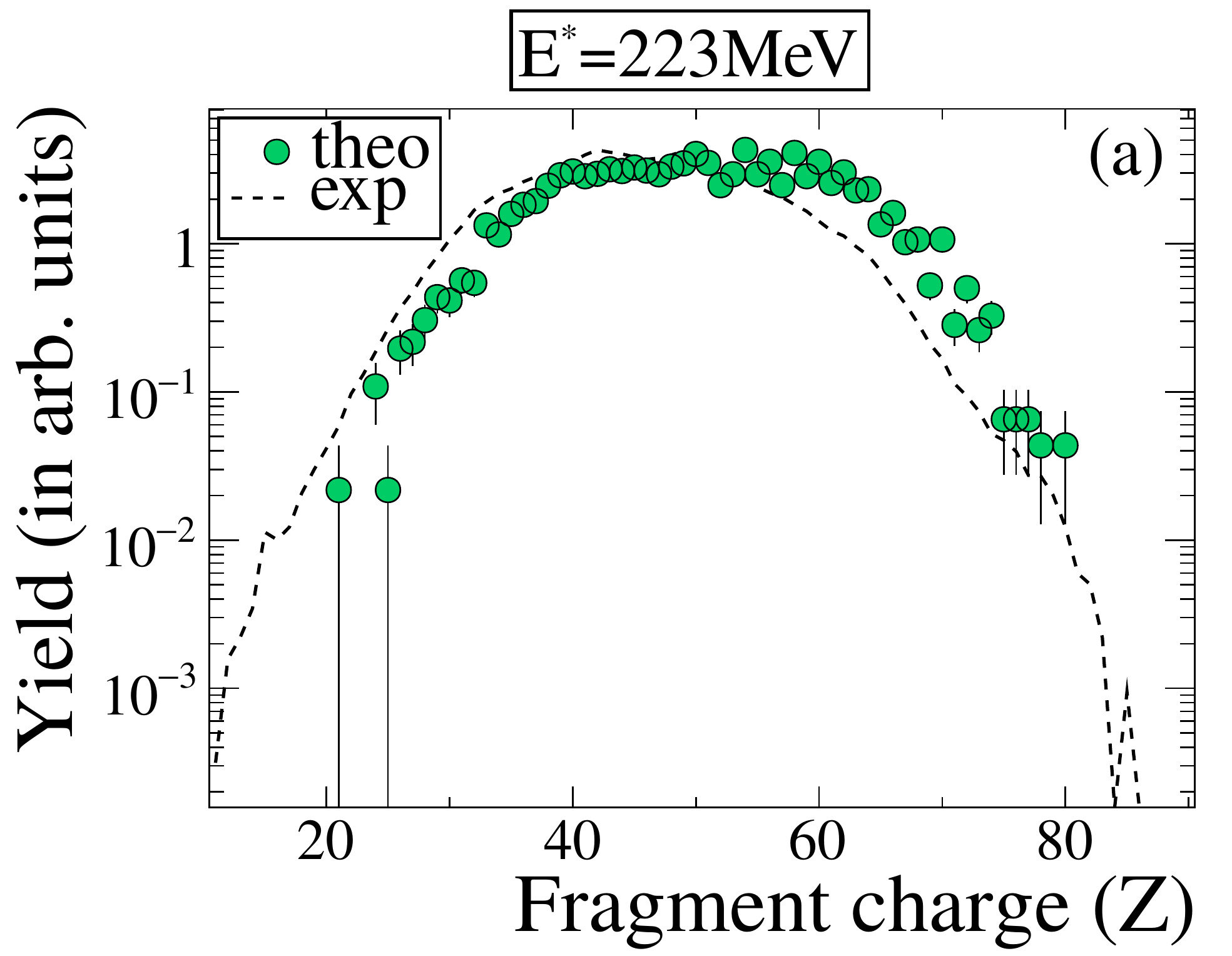}
\includegraphics[width=.32\textwidth]{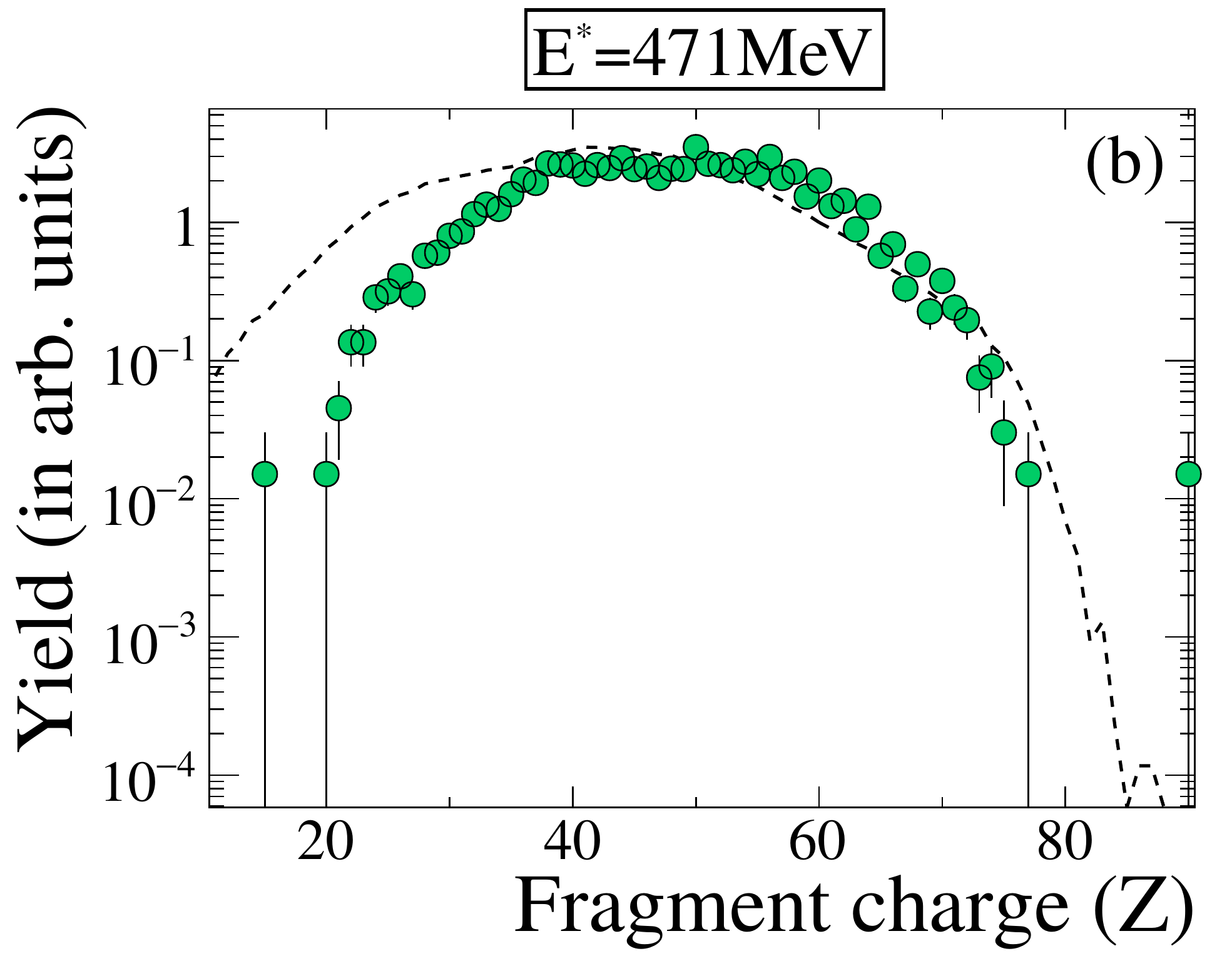}
\includegraphics[width=.339\textwidth]{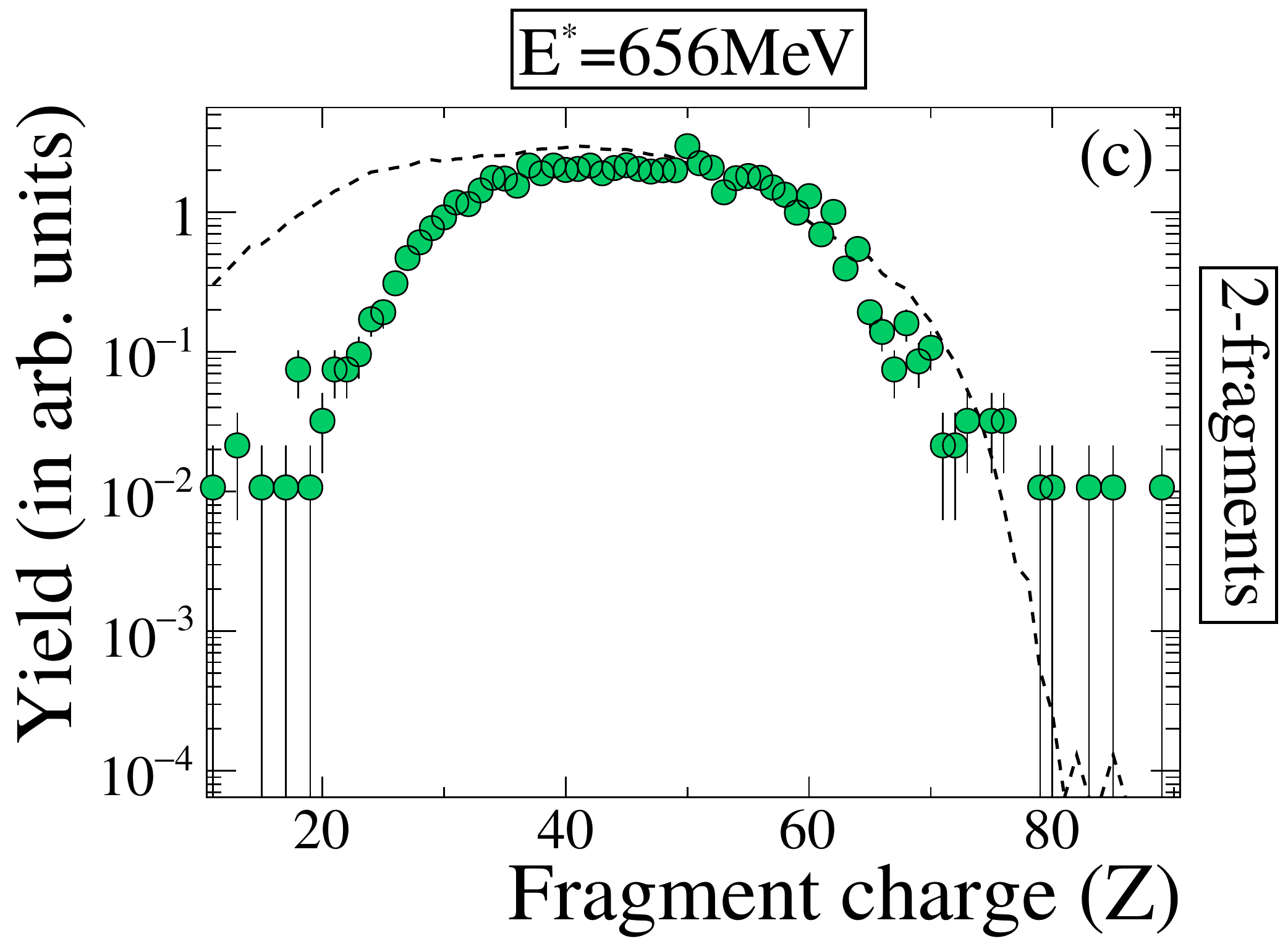}\\
\includegraphics[width=.32\textwidth]{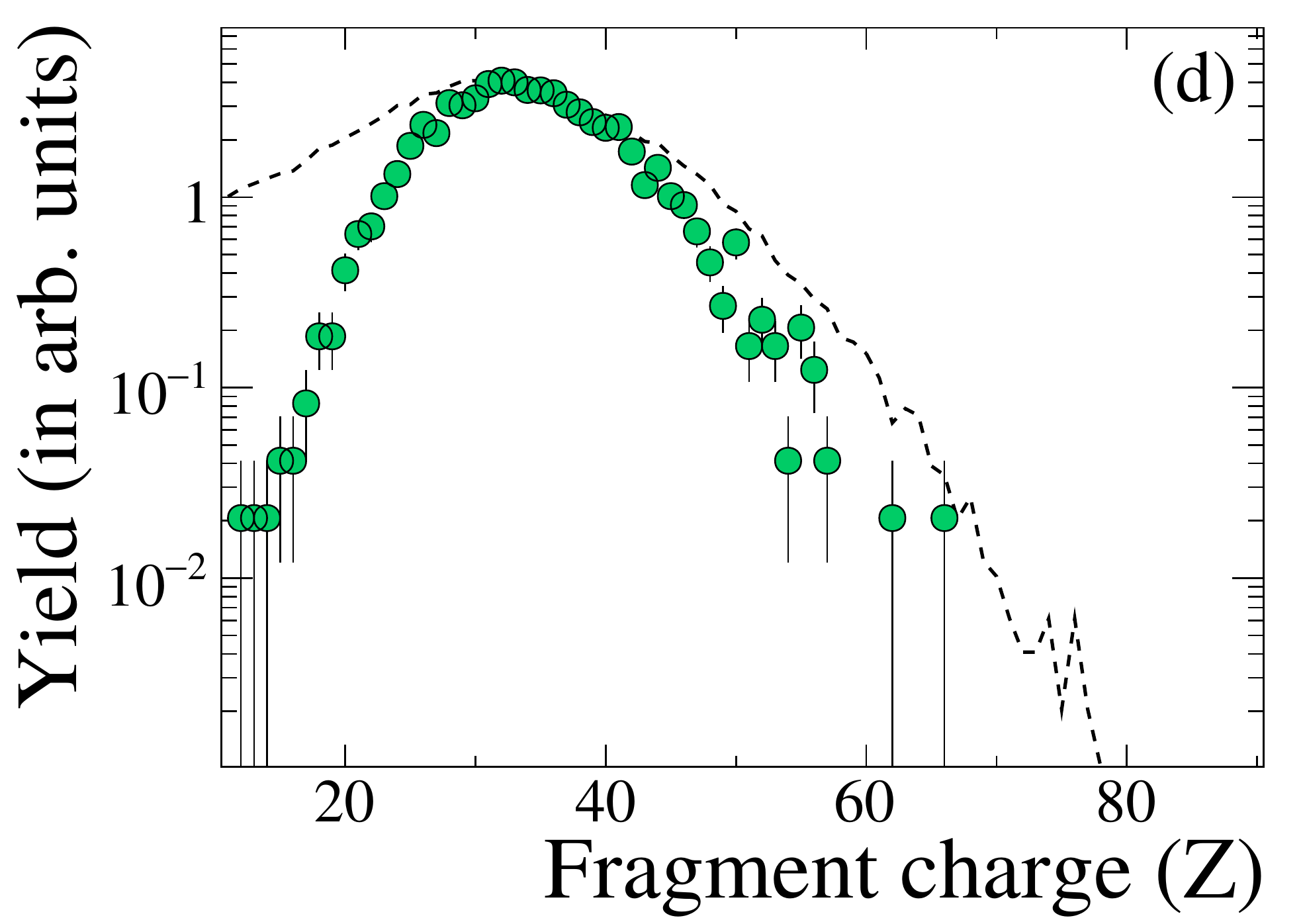}
\includegraphics[width=.32\textwidth]{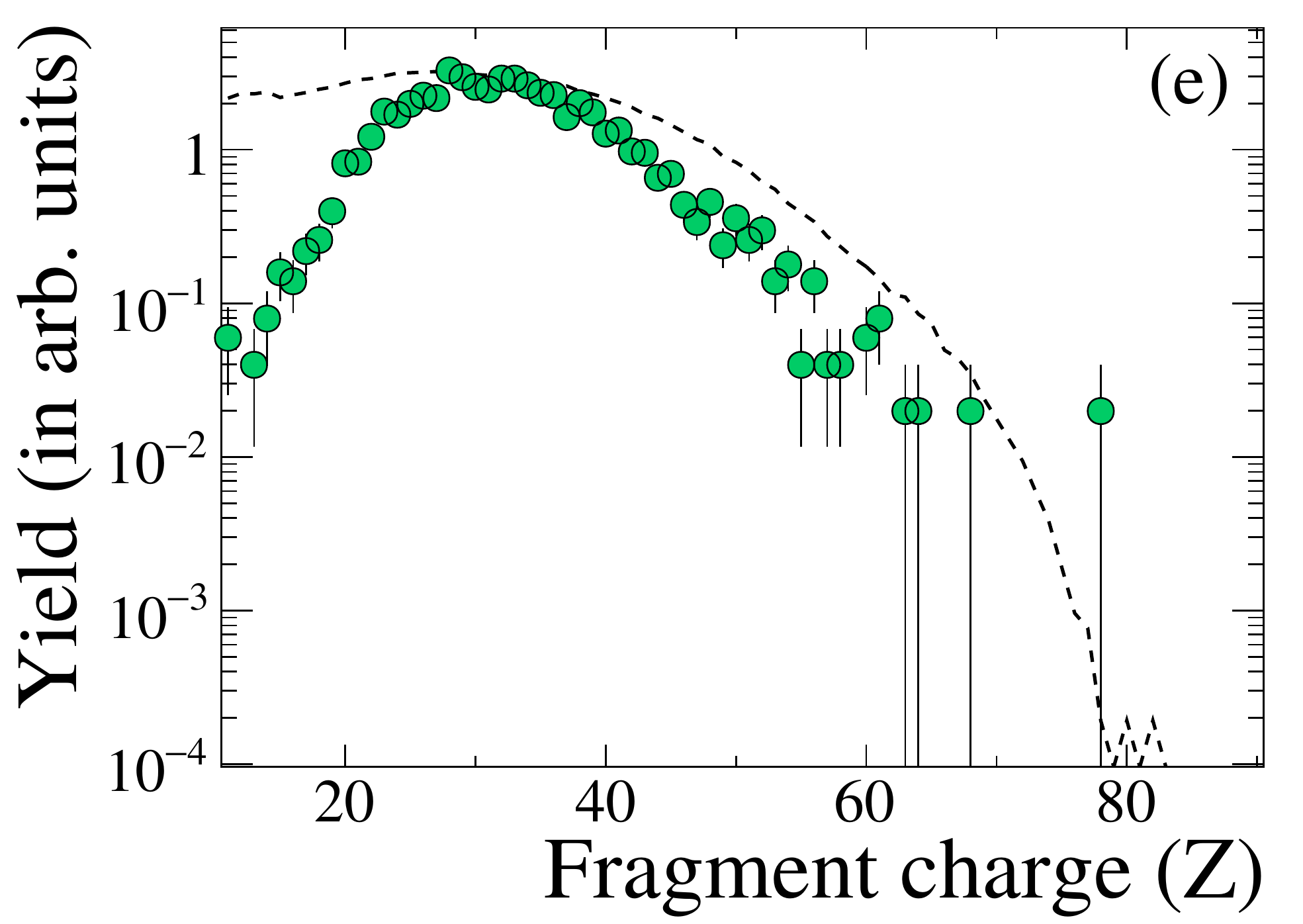}
\includegraphics[width=.339\textwidth]{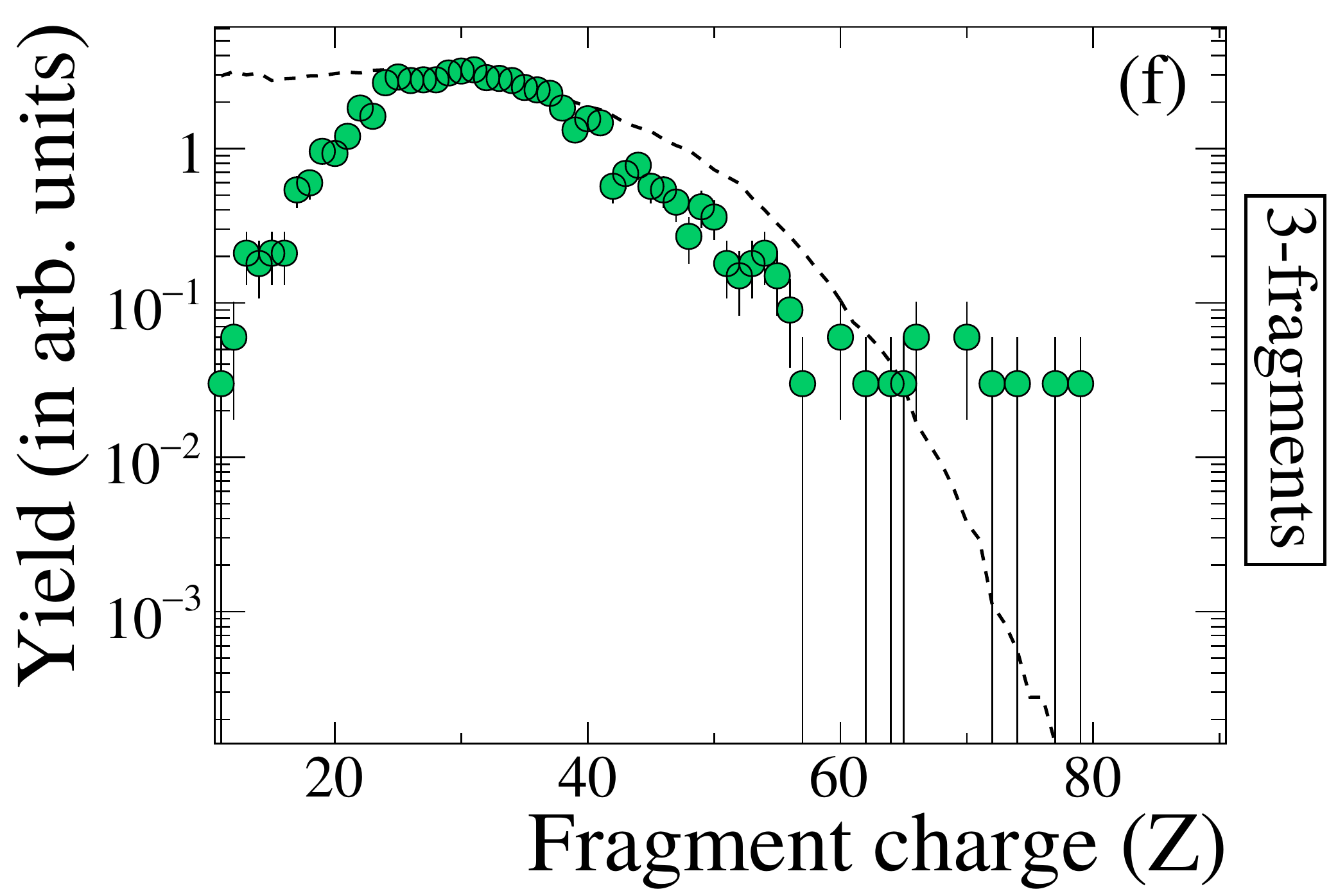}
\caption{(color online) The final fragment charge distributions for reaction $^{129}$Xe+$^{119}$Sn at excitation energies $E^*$= 223 (a,d), 471(b,e) and 656~MeV (c,f). 
The upper row corresponds to 2-fragment final state while the bottom shows 3-fragment events. 
The points present the theoretical calculations with its uncertainty and the lines are for experimental data. }
\label{zfinal_6}
\end{center}
\end{figure*}

The experimental data measured with INDRA are very complex but contain information about the richness of the processes occurring in this beam energy region. 
The \jdf{reasonable} reproduction of the total charge emitted as \jdf{light} particles \jdf{over a range of excitation energies} confirms the ability of the classical transport equations to describe the \jdf{underlying physics of the process}. 
 
\subsection{Charge distributions}
Fig.~\ref{zfinal_6} shows the charge distribution for the three excitation energies $E^*$= 223, 471 and 656~MeV 
(beam energies 8, 12 and 15 MeV/nucleon respectively) in columns. The upper row shows the results with the condition of producing two fragments with charge greater than 10 (M$_{\text{frag}}$=2) and 
the bottom row gives the events where one secondary fission occurres (M$_{\text{frag}}$=3). 

For the lowest excitation energy considered here ($E^*$= 223~MeV) with only two \jdf{final} fragments, the fragment charge distribution is well reproduced (Fig.~\ref{zfinal_6}(a)). 
This is \jdf{reassuring} since this system lies in the standard excitation energy region for the fission model. 
With increasing beam energy (Fig.~\ref{zfinal_6}(b-c)), the heavy fragment part of the distribution is still well reproduced. However, the model \jdf{underestimates the production of lighter fragments ($Z<40$)}.
\jdf{For 3-fragment events (Fig.~\ref{zfinal_6}(d-f)), where one of the primary fission fragments has undergone a secondary fission,
a resonable agreement is achieved concerning the distribution of the heaviest ($Z>30$) fragments.}
However, the same \jdf{underestimation} of low-Z fragments is observed, even for the lowest \jdf{excitation/beam} energy.

\jdf{In order to improve the agreement between the model and data we would need a more realistic description of the initial step of the reactions.}
\jdf{Indeed, the calculations presented here always assume the formation of a fully-equilibrated, spherical compound nucleus by complete fusion of $^{129}$Xe+$^{119}$Sn. Although this assumption seems to work quite well for the fission-like events at 8~MeV/nucleon (Fig.\ref{zfinal_6}(a)), it certainly cannot hold for all of the reactions presented here. Pre-equilibrium emission will become more important with increasing beam energy, leading to compound nuclei with a distribution of charge, mass and excitation energy. Also, prescission emission in the model is limited to neutrons, protons, deuteron, tritons and alpha particles, whereas at high excitation energies light fragments may also be evaporated. For such a heavy colliding system, the initial composite system formed in the reactions is likely to be strongly deformed rather than starting from a spherical shape, changing its subsequent dissipative trajectory in the PES.}

Preliminary calculations done with the microscopic code HIPSE \cite{lacroix:2004} shows that at such high beam energies the pre-equilibrium emission is very important:
during the time between the collision and the equilibration in form of the compound nucleus, some particles and also light fragments are emitted. 
\jdf{Thus the assumption of complete fusion in such an energy range as it is discussed presently is certainly not correct, and should be improved in future investigations. Our goal in this paper has been to establish the soundness of the principle that sequential fission of highly-excited nuclei, calculated for the first time with a realistic model of the fission process, can account at least qualitatively for the $N$-fragment exit channels observed experimentally.}

\section{Summary}
\label{conclu}

In this work we developed a model to describe 
\jdf{sequential fission of highly-excited compound nuclei}. The procedure can create
up to four final fragments treating the two fission steps in a coherent and fully dynamical way.
Primary fission of the compound nucleus is obtained by solving the coupled Langevin equations in a 4-dimensional collective coordinate landscape. After this first step,
fission fragments are considered as potential candidates for secondary fission and treated in the same way as the compound nucleus. Light particles as well as $\gamma$ ray's
are emitted statistically \jdf{all along} the deexcitation chain. This procedure allows to produce from one to four heavy fragments. 

The model was applied to the deexcitation of \jdf{a spherical} $^{248}$Rf nucleus with \jdf{excitation energies of} $E^*$= 223, 471, and 656~MeV corresponding to the complete fusion of 
$^{129}$Xe+$^{119}$Sn at 8, 12, and 15~MeV/nucleon. Model predictions were compared with experimental data on these reactions measured with the INDRA multi-detector. A satisfying agreement
was obtained showing the ability of the classical transport equations to describe deexcitation processes in this energy range.

Sequential fission could represent a unique tool to investigate the evolution of viscosity with temperature in a range of excitation energy inaccessible with standard fission models.
In particular, secondary fission probability appears to be very sensitive to viscosity/dissipation. However, before investigating different viscosity parametrizations,
entrance channel effects should be included. This will be done by coupling the present model with HIPSE calculation to simulate pre-equilibrium emissions.

{\bf{ Acknowledgements}}
\\
The authors express their acknowledgement to the INDRA collaboration for allowing them to use their partially unpublished data.
The work was partially supported by the Polish
National Science Centre under Contract No. 2013/08/M/ST2/00257 (LEA COPIGAL) (Project No.~18) and IN2P3-COPIN (Project No.~12-145, 09-146), and by the Russian Foundation for Basic Research (Project No.~13-02-00168). 

\bibliography{cites_prc}

\end{document}